\documentclass[aps,nofootinbib,superscriptaddress,floatfix]{revtex4} 
\usepackage{amsmath}
\usepackage{graphicx,epsfig,wrapfig}
\usepackage{amsthm}
\usepackage{makeidx}
\usepackage{amsfonts}
\usepackage{amssymb}
\usepackage{amsmath}
\usepackage{natbib}
\usepackage{dsfont}
\usepackage[capitalize]{cleveref}
\usepackage{mathrsfs}
\usepackage{slashed}
\usepackage{epstopdf}
\usepackage{bm}
\usepackage{color}
\usepackage[normalem]{ulem}
\usepackage{simplewick}
\renewcommand\sout{\bgroup \color[rgb]{0.55,0.00,0.99} \ULdepth=-.5ex \ULset}

\begin{document}

\newcommand{\ud}{\text{d}}
\newcommand{\mb}[1]{\boldsymbol{#1}}
\newcommand{\mc}[1]{\mathcal{#1}}
\newcommand{\non}{\nonumber}
\newcommand{\pep}[1]{\mathbf{#1}_{\perp}}
\newcommand{\pepgr}[1]{\bm{#1}_{\perp}}
\newcommand{\gdir}[1]{\gamma^{#1}}
\newcommand{\xk}{(x,\mathbf{k}_{\perp})}
\newcommand{\xkq}{(x,\mathbf{k}_{\perp}^{2})}
\newcommand{\pe}{(\mb{p}_{e})}
\newcommand{\xks}{(x,\pep{k};S)}
\newcommand{\pv}{\mathrm{P.V.}}
\newcommand{\xkl}{(x,\pep{k};\Lambda)}
\newcommand{\dEF}[1]{\nabla^{\text{EF}}_j\left(#1\right)}

\title{Spatial distribution of angular momentum inside the nucleon}

\author{C\'edric Lorc\'e}
\email{cedric.lorce@polytechnique.edu}
\affiliation{Centre de Physique Th\'eorique, \'Ecole polytechnique, 
	CNRS, Universit\'e Paris-Saclay, F-91128 Palaiseau, France}

\author{Luca Mantovani}
\email{luca.mantovani@pv.infn.it}
\affiliation{Dipartimento di Fisica, Universit\`a degli Studi di Pavia, I-27100 Pavia, Italy}
\affiliation{Istituto Nazionale di Fisica Nucleare, Sezione di
  Pavia,  I-27100 Pavia, Italy}

\author{Barbara Pasquini}
\email{barbara.pasquini@pv.infn.it}
\affiliation{Dipartimento di Fisica, Universit\`a degli Studi di Pavia, I-27100 Pavia, Italy}
\affiliation{Istituto Nazionale di Fisica Nucleare, Sezione di
  Pavia,  I-27100 Pavia, Italy}

\date{\today}

\allowdisplaybreaks[2]

\begin{abstract}
We discuss in detail the spatial distribution of angular momentum inside the nucleon. We show that the discrepancies between different definitions originate from terms that integrate to zero. Even though these terms can safely be dropped at the integrated level, they have to be taken into account at the density level. Using the scalar diquark model, we illustrate our results and, for the first time, check explicitly that the equivalence between kinetic and canonical orbital angular momentum persists at the density level, as expected in a system without gauge degrees of freedom.
\end{abstract}

\maketitle

\section{Introduction}\label{sec1}

Understanding how the spin of the nucleon originates from the spin and orbital motion of its constituent is one of the current key questions in hadronic physics. While this problem may seem rather straightforward in the context of ordinary quantum mechanics, it becomes quite challenging in the context of hadronic physics where one has to include relativistic, gauge-symmetry and non-perturbative aspects. One of the main conceptual issues is that the decomposition of the nucleon spin is not unique~\cite{Leader:2013jra,Wakamatsu:2014zza,Liu:2015xha}. This intrinsic ambiguity is sometimes considered as a sign indicating that the question is not physical. It actually reflects the fact that any decomposition necessarily relies on how one defines the degrees of freedom. The problem remains physical as long as the various contributions can in principle be accessed by experiments. 

Ji has shown that the (kinetic) total angular momentum of quarks and gluons can be expressed in terms of generalized parton distributions (GPDs)~\cite{Ji:1996ek}. This triggered an intense experimental program since GPDs can be extracted from exclusive processes like deeply virtual Compton scattering and hard meson exclusive electroproduction~\cite{Diehl:2003ny,Boffi:2007yc,Bacchetta:2016ccz}. Interestingly, the connection between GPDs and angular momentum has been clearly established only at the level of integrated quantities over all space. As shown by Burkardt~\cite{Burkardt:2000za,Burkardt:2002hr}, GPDs contain information about the spatial distribution of quarks and gluons inside the nucleon. It is therefore conceivable that GPDs contain also the information about the spatial distribution of angular momentum. The problem now is to determine how this information is precisely encoded.

Polyakov provided the first attempt to answer this question~\cite{Polyakov:2002yz}, but he required the nucleon to be infinitely massive, so as to avoid relativistic corrections. The infinite mass assumption can actually be relaxed, provided that one works within the light-front formalism, as sketched in the review~\cite{Leader:2013jra}. Recently, Adhikari and Burkardt compared different definitions of the angular momentum density and reached the conclusion that none of the definitions agree at the density level. They attributed some of the discrepancies to missing total divergence terms, as it had been pointed out earlier in Refs.~\cite{Leader:2013jra,Liu:2015xha}.

The purpose of the present paper is to revisit the work of Polyakov, discuss in more detail the alternative approach based on the light-front formalism, and identify all the missing terms that hinder the proper comparison of the various definitions of angular momentum. 

The rest of the paper is organized as follows. In \cref{sec2} we recall the connection between the energy-momentum tensor and angular momentum. We stress in particular that, unlike in General Relativity, the energy-momentum tensor is generally not symmetric in Particle Physics, owing to the presence of a spin density. In \cref{sec3} we derive three-dimensional densities of angular momentum in the Breit frame. We show that by projecting these densities onto a two-dimensional plane, they can be considered in the more general class of elastic frames. In \cref{sec4} we discuss the densities in the light-front formalism and observe that they coincide (for the longitudinal component of angular momentum) with the two-dimensional densities in the elastic frame. We illustrate our results within the scalar diquark model in \cref{sec5} and, for the first time, check explicitly that kinetic and canonical orbital angular momentum coincide at the density level in absence of gauge bosons. Finally, in \cref{sec6} we summarize our findings and draw our conclusions.

\section{Energy-momentum and generalized angular momentum tensors}\label{sec2}

In field theory, the conserved current associated with the invariance of the theory under Lorentz transformations, known as generalized angular momentum tensor, can be written in general as the sum of two contributions
\begin{equation}
J^{\mu\alpha\beta}(x)=L^{\mu\alpha\beta}(x)+S^{\mu\alpha\beta}(x) \; . \label{J}
\end{equation}
Each one of these tensors is antisymmetric under $\alpha\leftrightarrow\beta$. The first contribution reads
\begin{equation}
L^{\mu\alpha\beta}(x)= x^{\alpha}T^{\mu\beta}(x)-x^{\beta}T^{\mu\alpha}(x) \; ,
\end{equation}
where $T^{\mu\nu}(x)$ is the Energy-Momentum Tensor (EMT) density associated with the system, which accounts for the fact that the fields are affected by Lorentz transformations owing to their dependence on space-time points. The second contribution $S^{\mu\alpha\beta}(x)$ accounts for the fact that fields have in general many components, which can also be affected by Lorentz transformations. 

The three generators of rotations are obtained when $\alpha,\beta=i,j$ are spatial components. In this case, Eq.~\eqref{J} simply indicates that the total Angular Momentum (AM) is the sum of Orbital Angular Momentum (OAM) and spin
\begin{equation}
\mb{J}=\mb{L}+\mb{S}
\end{equation}
with $J^i=\frac{1}{2}\,\epsilon^{ijk} \int{\ud^3\boldsymbol{r}\,J^{0jk}}$, likewise for $L^i$ and $S^i$.

\subsection{Belinfante-improved tensors}

The energy-momentum tensor obtained by following the procedure in Noether's theorem is referred to as the canonical EMT, and is in general neither gauge invariant nor symmetric. Belinfante and Rosenfeld~\cite{Belinfante1, Belinfante2,Rosenfeld} proposed to add a so-called superpotential term to the definition of both the energy-momentum and generalized angular momentum tensors, defining the Belinfante-improved tensors as 
\begin{align}
T^{\mu\nu}_{\text{Bel}}(x)&= T^{\mu\nu}(x)+\partial_{\lambda}G^{\lambda\mu\nu}(x),\label{tbel} \\ 
J^{\mu\alpha\beta}_{\text{Bel}}(x)&= J^{\mu\alpha\beta}(x)+\partial_{\lambda}\!\left[x^{\alpha}G^{\lambda\mu\beta}(x)-x^{\beta}G^{\lambda\mu\alpha}(x)\right]\; , \label{jbel}
\end{align}
where  the superpotential $G^{\lambda\mu\nu}$ is given by the combination
\begin{equation}\label{G}
G^{\lambda\mu\nu}(x)=\frac{1}{2}\left[S^{\lambda\mu\nu}(x)+S^{\mu\nu\lambda}(x)+S^{\nu\mu\lambda}(x)\right]=-G^{\mu\lambda\nu}(x) \; .
\end{equation}
The effect of such a term is to modify the definition of the local density without changing the total charge.
The Belinfante-improved tensors \eqref{tbel}-\eqref{jbel} are conserved and usually turn out to be gauge invariant. Moreover, the particular choice~\eqref{G} allows us to write the total AM in a pure orbital form
\begin{equation}
J^{\mu\alpha\beta}_{\text{Bel}}(x)=x^{\alpha}T_{\text{Bel}}^{\mu\beta}(x)-x^{\beta}T_{\text{Bel}}^{\mu\alpha}(x).
\end{equation}
Since the new tensors are conserved, it follows from this expression that the Belinfante-improved EMT is symmetric.

\subsection{Kinetic tensors}

As discussed in Refs.~\cite{Leader:2013jra,Lorce:2015lna}, the requirement of a symmetric EMT is usually motivated by General Relativity. In that context, the notion of spin is not accounted for from the beginning, and it is natural to consider AM as purely orbital. From a Particle Physics perspective, however, one naturally includes a spin contribution to the total AM as in Eq.~\eqref{J}. It then follows from the conservation of both $T^{\mu\nu}(x)$ and $J^{\mu\alpha\beta}(x)$ that the EMT is in general asymmetric, the antisymmetric part being given by the divergence of the density of spin
\begin{equation}
T^{[\alpha\beta]}(x)=-\partial_{\mu}S^{\mu\alpha\beta}(x),  \label{asymm2}
\end{equation}
where $a^{[\mu}b^{\nu]}=a^\mu b^\nu- a^\nu b^\mu$.
We see the Belinfante-improved tensors as effective densities, where the effects of spin are mimicked by an obscure new contribution to momentum. Interestingly, recent developments in optics also seem to demote the Belinfante-improved expressions from their status as fundamental densities~\cite{Bliokh:2015doa}.

Instead of the Belinfante-improved tensors, Ji~\cite{Ji:1996ek} proposed to use in the context of QCD the kinetic EMT
\begin{equation}
T^{\mu\nu}_\text{kin}(x)=T^{\mu\nu}_{\text{kin},q}(x)+T^{\mu\nu}_{\text{kin},g}(x),
\end{equation}
where the gauge-invariant quark and gluon contributions are given by~\cite{Leader:2013jra,Lorce:2015lna}
\begin{align}
T^{\mu\nu}_{\text{kin},q}(x)&=\frac{1}{2}\,\overline{\psi}(x)\gamma^{\mu}i\overleftrightarrow{D}^{\nu}\psi(x) \;,\\
T^{\mu\nu}_{\text{kin},g}(x)&=-2\,\text{Tr}\left[G^{\mu\lambda}(x)G^\nu_{\phantom{\nu}\lambda}(x)\right]+\frac{1}{2}\,g^{\mu\nu}\,\text{Tr}[G^{\rho\sigma}(x)G_{\rho\sigma}(x)] \; ,
\end{align}
with $\overleftrightarrow{D}^{\mu}=\overleftrightarrow{\partial}^{\mu}-igA^{\mu}$ and $\overleftrightarrow{\partial}^{\mu}=\overrightarrow{\partial}^{\mu}-\overleftarrow{\partial}^{\mu}$, and the field-strength tensor $G_{\mu\nu}(x)=\partial_\mu A_\nu(x)-\partial_\nu A_\mu(x)-ig\left[A_\mu(x),A_\nu(x)\right]$. The kinetic generalized AM tensor reads
\begin{equation}
J^{\mu\alpha\beta}_\text{kin}(x)=L^{\mu\alpha\beta}_{\text{kin},q}(x)+S^{\mu\alpha\beta}_{q}(x)+J^{\mu\alpha\beta}_{\text{kin},g}(x)
\end{equation}
with 
\begin{align}
L^{\mu\alpha\beta}_{\text{kin},q}(x)&=x^\alpha T^{\mu\beta}_{\text{kin},q}(x)-x^\beta T^{\mu\alpha}_{\text{kin},q}(x) \; ,\\
S^{\mu\alpha\beta}_{q}(x)&=\frac{1}{2}\,\varepsilon^{\mu\alpha\beta\lambda}\,\overline{\psi}(x)\gamma_{\lambda}\gamma_{5}\psi(x) \; ,\label{Sq}\\
J^{\mu\alpha\beta}_{\text{kin},g}(x)&=x^\alpha T^{\mu\beta}_{\text{kin},g}(x)-x^\beta T^{\mu\alpha}_{\text{kin},g}(x) \; .
\end{align}
and the convention $\varepsilon_{0123}=+1$. Contrary to the quark total AM, the gluon total AM cannot be split into orbital and spin contributions which are at the same time gauge-invariant and local~\cite{Lorce:2012rr,Lorce:2013bja}.
The kinetic and Belinfante-improved tensors in QCD are related as follows
\begin{align}
T^{\mu\nu}_{\text{kin},q}(x)&= T^{\mu\nu}_{\text{Bel},q}(x)-\frac{1}{2}\,\partial_{\lambda}S^{\lambda\mu\nu}_q(x) \; ,\\
L^{\mu\alpha\beta}_{\text{kin},q}(x)+S^{\mu\alpha\beta}_q(x)&= J^{\mu\alpha\beta}_{\text{Bel},q}(x)-\frac{1}{2}\,\partial_{\lambda}\!\left[x^{\alpha}S^{\lambda\mu\beta}_q(x)-x^{\beta}S^{\lambda\mu\alpha}_q(x)\right] \; ,\label{relation}
\end{align}
the gluon contributions being the same in both cases, $T^{\mu\nu}_{\text{kin},g}(x)= T^{\mu\nu}_{\text{Bel},g}(x)$ and $J^{\mu\alpha\beta}_{\text{kin},g}(x)=J^{\mu\alpha\beta}_{\text{Bel},g}(x)$. Using the conservation of the total AM $J^{\mu\alpha\beta}_{\text{kin}}$ and the symmetry of $T^{\mu\nu}_{\text{kin},g}(x)$, one can relate the antisymmetric part of the quark kinetic EMT to the quark spin divergence
\begin{equation}
T^{[\alpha\beta]}_{\text{kin},q}(x)=-\partial_\mu S^{\mu\alpha\beta}_q(x), 
\end{equation}
or more explicitly
\begin{equation}
\overline{\psi}(x)\gamma^{[\alpha}i\overleftrightarrow{D}^{\beta]}\psi(x)=-\varepsilon^{\alpha\beta\mu\lambda}\,\partial_{\mu}\!\left[\overline{\psi}(x)\gamma_{\lambda}\gamma_{5}\psi(x)\right] \; , \label{asymm} 
\end{equation}
as one can also derive directly from the QCD equations of motion.
It then follows that the Belinfante-improved EMT just coincides with the symmetric part of the kinetic EMT
\begin{equation}\label{kinbel}
\frac{1}{2}\,T^{\{\mu\nu\}}_{\text{kin},a}(x)=T^{\mu\nu}_{\text{Bel},a}(x)\; ,\qquad\qquad a=q,g
\end{equation}
where $a^{\{\mu}b^{\nu\}}=a^\mu b^\nu+ a^\nu b^\mu$. This simple relation holds only owing to the total antisymmetry of the spin contribution. \\
Since kinetic and Belinfante-improved tensors differ by superpotential terms, they lead to the same charges. For this reason, the superpotentials are often dropped from the discussions in the literature. However, once one goes back to the density level, it is crucial to pay attention to these terms.

\subsection{Parametrization in terms of form factors}

We are interested in the matrix elements of the above-mentioned density operators. It will be sufficient to consider the operators evaluated at $x=0$, since the general case is recovered simply through a translation of fields. Moreover, since the average position is the Fourier conjugate variable to the momentum transfer $\Delta$, we need to consider off-forward matrix elements. As shown by Bakker, Leader and Trueman~\cite{Bakker:2004ib}, the matrix elements of the general local asymmetric energy-momentum tensor for a spin-$1/2$ target are parametrized in terms of five form factors~\cite{Leader:2013jra}:
\begin{align}
\langle p',\mb s'\lvert T^{\mu\nu}(0) \rvert p,\mb s\rangle& = \overline{u}(p',\mb s')\left[\frac{P^{\mu}P^\nu}{M}\,A(t)+\frac{P^{\mu}i\sigma^{\nu\lambda}\Delta_{\lambda}}{4M}\,(A+B+D)(t)\right. \non \\
&\left.\qquad+\frac{\Delta^{\mu}\Delta^{\nu}-g^{\mu\nu}\Delta^2}{M}\,C(t) +Mg^{\mu\nu}\,\bar{C}(t)+\frac{P^{\nu}i\sigma^{\mu\lambda}\Delta_{\lambda}}{4M}\,(A+B-D)(t)\right]u(p,\mb s) \; , \label{dec4}
\end{align} 
where $M$ is the nucleon mass, the three-vectors $\mb s$ and $\mb{s'}$ (with $\mb{s}^2=\mb{s'}^2=1$) denote the rest-frame polarization of the initial and final states, respectively, and
\begin{equation}
P=\frac{p'+p}{2}, \qquad \Delta=p'-p ,\qquad t=\Delta^2 .
\end{equation}
The onshell conditions for initial and final states $p^2=p'^2=M^2$ are equivalent to
\begin{equation}
P^2=M^2-\frac{\Delta^2}{4}, \qquad P\cdot\Delta=0. \label{a}
\end{equation}

Beside the EMT, we also need a parametrization of the matrix elements of the quark spin operator $S^{\mu\alpha\beta}_q(0)$. Owing to Eq.~\eqref{Sq}, we can write
\begin{equation} 
\langle p',\mb s'\lvert S^{\mu\alpha\beta}_q(0)\rvert p,\mb s\rangle =\frac{1}{2}\,\varepsilon^{\mu\alpha\beta\lambda}\,\overline{u}(p',\mb s')\left[\gamma_{\lambda}\gamma_5\, G^q_{A}(t)+\frac{\Delta_{\lambda}\gamma_5}{2M}\,G^q_{P}(t)\right]u(p,\mb s), \label{spar}
\end{equation}
where $G^q_{A}(t)$ and $G^q_{P}(t)$ are the axial-vector and induced pseudoscalar form factors, respectively. It then follows from the QCD identity~\eqref{asymm} that the form factor associated with the antisymmetric part of the quark EMT is related to the axial-vector form factor~\cite{Bakker:2004ib,Leader:2013jra}
\begin{equation}
D_q(t)=-G^q_A(t)\; .
\end{equation}

\section{Densities in instant form}\label{sec3}

Inspired by Sachs' interpretation of the electromagnetic form factors in the Breit frame~\cite{Sachs:1962zzc}, Polyakov and collaborators discussed the spatial distribution of angular momentum in instant form based on the Belinfante form of the EMT~\cite{Polyakov:2002yz,Polyakov:2002wz,Goeke:2007fp,Cebulla:2007ei}. We revisit this discussion in more detail, using this time the more general asymmetric EMT. From now on we drop the label ``kin'' in all kinetic quantities, as well as the reference to quarks and gluons.

\subsection{3D densities in the Breit frame}\label{3dbf}

Let us start with the definition of kinetic OAM distribution in four-dimensional position space
\begin{equation}
\langle L^i\rangle(x)=\varepsilon^{ijk}\,x^j\int\frac{\ud^{3}\boldsymbol{\Delta}}{(2\pi)^3\,2\sqrt{p'^0p^0}}\,\langle p',\mb s\lvert T^{0k}(x)\rvert p,\mb s\rangle= \varepsilon^{ijk}\,x^j\int\frac{\ud^{3}\boldsymbol{\Delta}}{(2\pi)^3}\,e^{i\Delta\cdot x}\,\langle T^{0k}\rangle, \label{l1} 
\end{equation}
where we introduced for convenience
\begin{equation}
\langle T^{\mu\nu}\rangle\equiv\frac{\langle p',\mb s\lvert T^{\mu\nu}(0)\rvert p,\mb s\rangle}{2\sqrt{p'^0p^0}} \;.\label{short} 
\end{equation}
Notice that the energy transfer $\Delta^{0}$ is not an independent variable but a function of the three-momentum transfer $\mb{\Delta}$ through the onshell conditions~\eqref{a}
\begin{equation} 
\Delta^{0}=\frac{\mb{P}\cdot\mb{\Delta}}{P^0},\qquad P^0=\frac{1}{2}\left[\sqrt{\left(\mb P+\frac{\mb \Delta}{2}\right)^2+M^2}+\sqrt{\left(\mb P-\frac{\mb \Delta}{2}\right)^2+M^2}\right] . \label{d0} 
\end{equation}
Using integration by parts, and disregarding as usual the surface term, we rewrite Eq.~\eqref{l1} as
\begin{equation} 
\langle L^i\rangle(x)=\varepsilon^{ijk}\int\frac{\ud^{3}\boldsymbol{\Delta}}{(2\pi)^3}\,e^{i\Delta\cdot x}\left[-i\frac{\partial\langle T^{0k}\rangle}{\partial\Delta^{j}}+\frac{x^{0}}{2}\left(\frac{p'^j}{p'^0}+\frac{p^j}{p^0}\right)\langle T^{0k}\rangle\right]. \label{lix}
\end{equation}
The second term is in general different from zero. Its explicit time dependence comes from the non-conservation of the individual contributions to the total AM of the system. \\
One way to get rid of this term, along with the $x^0$ dependence in Eq.~\eqref{lix}, is to restrict ourselves to the Breit (or ``brick-wall'') frame (BF), defined by the condition $\mb{P}=\mb{0}$. This implies in particular $\Delta^0=0$ and $P^0=\sqrt{\frac{\mb\Delta^2}{4}+M^2}$. We can then define the spatial density of kinetic OAM as\footnote{We note in passing that the incorrect sign for the Fourier transform was used in~\cite{Polyakov:2002yz,Polyakov:2002wz,Goeke:2007fp,Cebulla:2007ei}.}
\begin{equation} 
\langle L^{i}\rangle(\mb{x})=-i\varepsilon^{ijk}\int\frac{\ud^{3}\boldsymbol{\Delta}}{(2\pi)^3}\,e^{-i\mb{\Delta}\cdot\mb{x}}\left.\frac{\partial\langle T^{0k}\rangle}{\partial\Delta^{j}}\right|_\text{BF}.  \label{nolim}
\end{equation}
This is indeed consistent with a density interpretation since $\mb p'=-\mb p$ implies that the initial and final wave functions undergo the same Lorentz contraction. \\
Using the general parametrization~\eqref{dec4} and taking the same rest-frame polarization three-vector $\mb s$ for both the initial and final states, we find that the kinetic OAM density reads (see Appendix \ref{a1} for more details)
\begin{equation}
\langle L^i\rangle(\mb{x})=\int\frac{\ud^3\boldsymbol{\Delta}}{(2\pi)^3}\,e^{-i\mb{\Delta}\cdot\mb{x}}\left[s^i\, L(t)+\left[(\mb{\Delta}\cdot\mb{s})\Delta^i-\mb{\Delta}^2s^i\right]\frac{\ud L(t)}{\ud t}\right]_{t=-\mb \Delta^2}, \label{lir2}
\end{equation}
where we introduced for convenience the combination of energy-momentum form factors
\begin{equation}
L(t)=\frac{1}{2}\left[A(t)+B(t)+D(t)\right].
\end{equation}
Similarly, for the spin density we find that
\begin{align}
\langle S^{i}\rangle(\mb{x})&=\frac{1}{2}\,\varepsilon^{ijk}\int\frac{\ud^{3}\boldsymbol{\Delta}}{(2\pi)^3}\,e^{-i\mb{\Delta}\cdot\mb{x}}\left.\langle S^{0jk}\rangle\right|_\text{BF}\nonumber\\
&=\int\frac{\ud^{3}\boldsymbol{\Delta}}{(2\pi)^3}\,e^{-i\mb{\Delta}\cdot\mb{x}}\,\left[\frac{s^i}{2}\, G_{A}(t)-\frac{\left(\mb{\Delta}\cdot\mb{s}\right)\Delta^i}{4}\,\frac{\ud G(t)}{\ud t}\right]_{t=-\mb \Delta^2}, \label{six}
\end{align}
where we introduced for convenience
\begin{equation}
\frac{\ud G(t)}{\ud t}=\frac{1}{2P^0}\left[\frac{G_A(t)}{P^0+M}+\frac{G_P(t)}{M}\right].
\end{equation}

Polyakov and collaborators~\cite{Polyakov:2002yz,Polyakov:2002wz,Goeke:2007fp,Cebulla:2007ei} considered the Belinfante-improved form of the EMT. Recalling that $T^{\mu\nu}_{\text{Bel}}=\frac{1}{2}T^{\{\mu\nu\}}$, it is easy to see that the density of Belinfante-improved total AM assumes the same structure as in Eq.~\eqref{lir2}, but now without the $D(t)$ contribution
\begin{equation}\label{Jbeldens}
\langle J^i_\text{Bel}\rangle(\mb{x})=\int\frac{\ud^3\boldsymbol{\Delta}}{(2\pi)^3}\,e^{-i\mb{\Delta}\cdot\mb{x}}\left[s^i\, J(t)+\left[(\mb{\Delta}\cdot\mb{s})\Delta^i-\mb{\Delta}^2s^i\right]\frac{\ud J(t)}{\ud t}\right]_{t=-\mb \Delta^2},
\end{equation}
where we used Polyakov's form factor
\begin{equation}
J(t)=\frac{1}{2}\left[A(t)+B(t)\right]. \label{PolJ}
\end{equation}
We can compare this expression with the kinetic total AM density $\left\langle J^{i}\right\rangle(\bm{x})=\left\langle L^{i}\right\rangle(\bm{x})+\left\langle S^{i}\right\rangle(\bm{x})$. From Eqs.~\eqref{lir2} and \eqref{six} and taking into account that $D(t)=-G_A(t)$, we find
\begin{equation}
\left\langle J^{i}\right\rangle (\bm{x})=\int\frac{\ud^{3}\boldsymbol{\Delta}}{(2\pi)^3}\,e^{-i\mb{\Delta}\cdot\mb{x}}\,\left[s^i\,J(t)+\left[(\mb{\Delta}\cdot\mb{s})\Delta^i-\mb{\Delta}^2s^i\right]\frac{\ud L(t)}{\ud t}-\frac{\left(\mb{\Delta}\cdot\mb{s}\right)\Delta^i}{4}\,\frac{\ud G(t)}{\ud t}\right]_{t=-\mb \Delta^2} \; . \label{jix}
\end{equation}
Therefore we have at the density level
\begin{equation}
\langle J^i\rangle(\mb{x})\neq\langle J^i_\text{Bel}\rangle(\mb{x}) \, ,
\end{equation}
while
\begin{equation}
\left\langle J^i\right\rangle =\int\ud^3\boldsymbol{x}\,\langle J^i\rangle(\mb{x})=\int\ud^3\boldsymbol{x}\,\langle J^i_\text{Bel}\rangle(\mb{x})=s^i J(0)\; , \label{ji}
\end{equation}
which is nothing but th Ji relation \cite{Ji:1996ek} in the rest frame of the target.
The reason for this mismatch is the total divergence in Eq.~\eqref{relation}. We obtain for the corresponding density
\begin{align}
\langle M^{i}\rangle(\mb{x})&=\frac{1}{2}\,\varepsilon^{ijk}\int\frac{\ud^{3}\boldsymbol{\Delta}}{(2\pi)^3}\,e^{-i\mb{\Delta}\cdot\mb{x}}\,\Delta^l\left.\frac{\partial\langle S^{l0k}\rangle}{\partial\Delta^j}\right|_\text{BF}\nonumber\\
&=-\int\frac{\ud^{3}\boldsymbol{\Delta}}{(2\pi)^3}\,e^{-i\mb{\Delta}\cdot\mb{x}}\left[\frac{\left[(\mb{\Delta}\cdot\mb{s})\Delta^i-\mb{\Delta}^2s^i\right]}{2}\,\frac{\ud G_{A}(t)}{\ud t}+\frac{\left(\mb{\Delta}\cdot\mb{s}\right)\Delta^i}{4}\frac{\ud G(t)}{\ud t}\right]_{t=-\mb \Delta^2}, \label{cix2}
\end{align}
leading then to
\begin{equation}
\langle J^i\rangle(\mb{x})=\langle J^i_\text{Bel}\rangle(\mb{x})+\langle M^{i}\rangle(\mb{x}) \; ,
\end{equation}
as expected.\\
Notice that, since integrating over $\bm x$ is equivalent to setting $\bm{\Delta}=\bm{0}$, from Eqs.~\eqref{Jbeldens}, \eqref{jix} and \eqref{ji} we can also write
\begin{equation}
J^i=s^i\int \ud^3\boldsymbol{x}\int\frac{\ud^{3}\boldsymbol{\Delta}}{(2\pi)^3}\,e^{-i\mb{\Delta}\cdot\mb{x}}\,J(-\bm{\Delta}^2) \; . \label{naive}
\end{equation}
It may therefore be tempting to interpret the Fourier transform of the form factor $J(t)$ as the density of total angular momentum. We see however from Eqs.~\eqref{Jbeldens} and \eqref{jix} that, in both the Belinfante's and in the kinetic case, other terms explicitly depending on $\bm{\Delta}$ do also contribute at the density level. More precisely, for the kinetic total AM we can introduce the following decomposition:
\begin{equation}
\left\langle J^i\right\rangle(\mb{x})=\left\langle J^i\right\rangle_{\text{naive}}(\mb{x})+\left\langle J^i\right\rangle_{\text{corr}}(\mb{x})
\end{equation}
into a ``naive'' contribution
\begin{equation}
\left\langle J^i\right\rangle_{\text{naive}}(\mb{x})=\int\frac{\ud^{3}\boldsymbol{\Delta}}{(2\pi)^3}\,e^{-i\mb{\Delta}\cdot\mb{x}}\,s^iJ(-\bm{\Delta^2}) \;  \label{jnaive}
\end{equation}
and a correction
\begin{equation}
\left\langle J^i\right\rangle_{\text{corr}}(\mb{x})= \int\frac{\ud^{3}\boldsymbol{\Delta}}{(2\pi)^3}\,e^{-i\mb{\Delta}\cdot\mb{x}}\,\left[\left[(\mb{\Delta}\cdot\mb{s})\Delta^i-\mb{\Delta}^2s^i\right]\frac{\ud L(t)}{\ud t}-\frac{\left(\mb{\Delta}\cdot\mb{s}\right)\Delta^i}{4}\,\frac{\ud G(t)}{\ud t}\right]_{t=-\mb \Delta^2} \; , \label{jcorr}
\end{equation}
satisfying
\begin{equation}
\int\ud^3\boldsymbol{x}\left\langle J^i\right\rangle_{\text{naive}}(\mb{x})=\left\langle J^i\right\rangle \; , \qquad \int\ud^3\boldsymbol{x}\left\langle J^i\right\rangle_{\text{corr}}(\mb{x})=0\; .
\end{equation}
Finally, in order to establish the connection with the results of~\cite{Polyakov:2002yz,Polyakov:2002wz,Goeke:2007fp,Cebulla:2007ei}, we decompose Eq.~\eqref{Jbeldens}
\begin{equation}
\langle J^i_\text{Bel}\rangle(\mb{x})=\langle J^i_\text{Bel}\rangle_{\text{mono}}(\mb{x})+\langle J^i_\text{Bel}\rangle_{\text{quad}}(\mb{x}),
\end{equation}
into monopole and quadrupole contributions
\begin{align}
\langle J^i_\text{Bel}\rangle_{\text{mono}}(\mb{x})&=s^i\int\frac{\ud^3 \boldsymbol{\Delta}}{(2\pi)^3}\,e^{-i\mb{\Delta}\cdot\mb{x}}\left[J(t)+\frac{2t}{3}\,\frac{\ud J(t)}{\ud t}\right]_{t=-\mb \Delta^2} , \label{lpol}\\
\langle J^i_\text{Bel}\rangle_{\text{quad}}(\mb{x})&=s^j\int\frac{\ud^3 \boldsymbol{\Delta}}{(2\pi)^3}\,e^{-i\mb{\Delta}\cdot\mb{x}}\,\left[\Delta^i\Delta^j-\frac{1}{3}\,\delta^{ij}\mb{\Delta}^2\right]\left.\frac{\ud J(t)}{\ud t}\right|_{t=-\mb \Delta^2}. \label{lquad}
\end{align}
The monopole contribution is the expression used by Polyakov and collaborators. As explained in the Appendix H of~\cite{Goeke:2007fp}, they discarded the quadrupole contribution because they interpreted it as an artifact originating from the non-covariance of the light-front formalism. Here we clearly see that the quadrupole term has actually nothing to do with the light-front formalism and simply arises from the breaking of spherical symmetry down to axial symmetry due to the polarization of the state. In conclusion, although the quadrupole term does not contribute once integrated over all space, it cannot be discarded when we consider densities. \\

\subsection{2D densities in the elastic frame}

The Breit frame allows one to define 3D densities for $\mb P=\mb 0$. If we want to consider the case where $\mb P\neq\mb 0$, the only densities we can define are necessarily two-dimensional. Indeed, in order to preserve the condition $\Delta^0=0$ which ensures that both the initial and final states are affected by the same Lorentz contraction factor, we have to restrict $\mb\Delta$ to the subspace orthogonal to $\mb P$. 

We define the \emph{elastic frames} (EF) by the condition $\mb P\cdot\mb\Delta=0$. They constitute a class of frames characterized by the fact that there is no energy transferred to the system, i.e.~$\Delta^0=0$; the energy of the system is then given by $P^0=\sqrt{\mb P^2+\frac{\mb\Delta^2}{4}+M^2}$. The Breit frame appears as a particular element of this class. 

Since $\mb P$ distinguishes a particular spatial direction, it is convenient to write three-vectors in terms of longitudinal and transverse components. Without loss of generality, we choose the spatial axes so that $\mb P$ lies along the $z$ axis
\begin{equation}
\mb P=(\mb 0_\perp,P),\qquad \mb\Delta=(\mb \Delta_\perp,0).
\end{equation}
In order to get rid of the time dependence in Eq.~\eqref{lix}, we will restrict ourselves to the longitudinal component of angular momentum only.
To comply with standard notations~\cite{Burkardt:2000za}, we will denote the Fourier conjugate variable to $\mb\Delta_\perp$ by $\mb b_\perp$ instead of $\mb x_\perp$. We then define the impact-parameter densities of kinetic OAM and spin as\begin{align}
\langle L^{z}\rangle(\mb{b}_\perp)&=-i\varepsilon^{3jk}\int\frac{\ud^{2}\boldsymbol{\Delta}_\perp}{(2\pi)^2}\,e^{-i\mb{\Delta}_\perp\cdot\mb{b}_\perp}\left.\frac{\partial\langle T^{0k}\rangle}{\partial\Delta^{j}_\perp}\right|_\text{EF}\nonumber\\
&=s^z\int\frac{\ud^2\boldsymbol{\Delta}_\perp}{(2\pi)^2}\,e^{-i\mb{\Delta}_\perp\cdot\mb{b}_\perp}\left[L(t)+t\,\frac{\ud L(t)}{\ud t}\right]_{t=-\mb \Delta^2_\perp} \; ,\label{L2D}\\
\langle S^{z}\rangle(\mb{b}_\perp)&=\frac{1}{2}\,\varepsilon^{3jk}\int\frac{\ud^{2}\boldsymbol{\Delta}_\perp}{(2\pi)^2}\,e^{-i\mb{\Delta}_\perp\cdot\mb{b}_\perp}\left.\langle S^{0jk}\rangle\right|_\text{EF}\nonumber\\
&=\frac{s^z}{2}\int\frac{\ud^{2}\boldsymbol{\Delta}_\perp}{(2\pi)^2}\,e^{-i\mb{\Delta}_\perp\cdot\mb{b}_\perp}\,G_{A}(-\mb \Delta^2_\perp) \; .
\end{align}
Similarly, for the impact-parameter densities of Belinfante-improved total AM and total divergence, we find
\begin{align}
\langle J^{z}_\text{Bel}\rangle(\mb{b}_\perp)&=-i\varepsilon^{3jk}\int\frac{\ud^{2}\boldsymbol{\Delta}_\perp}{(2\pi)^2}\,e^{-i\mb{\Delta}_\perp\cdot\mb{b}_\perp}\left.\frac{\partial\langle T^{0k}_\text{Bel}\rangle}{\partial\Delta^{j}_\perp}\right|_\text{EF}\nonumber\\
&=s^z\int\frac{\ud^2\boldsymbol{\Delta}_\perp}{(2\pi)^2}\,e^{-i\mb{\Delta}_\perp\cdot\mb{b}_\perp}\left[J(t)+t\,\frac{\ud J(t)}{\ud t}\right]_{t=-\mb \Delta^2_\perp},\\
\langle M^{z}\rangle(\mb{b}_\perp)&=\frac{1}{2}\,\varepsilon^{3jk}\int\frac{\ud^{2}\boldsymbol{\Delta}_\perp}{(2\pi)^2}\,e^{-i\mb{\Delta}_\perp\cdot\mb{b}_\perp}\,\Delta^l_\perp\left.\frac{\partial\langle S^{l0k}\rangle}{\partial\Delta^j_\perp}\right|_\text{EF}\nonumber\\
&=-\frac{s^z}{2}\int\frac{\ud^{2}\boldsymbol{\Delta}_\perp}{(2\pi)^2}\,e^{-i\mb{\Delta}_\perp\cdot\mb{b}_\perp}\left[t\,\frac{\ud G_{A}(t)}{\ud t}\right]_{t=-\mb \Delta^2_\perp}. \label{M2D}
\end{align}
The 2D distributions~\eqref{L2D}-\eqref{M2D} are axially symmetric and, remarkably, appear to be independent of $\mb P$. The reason is that longitudinal boosts do not mix longitudinal components of angular momentum. As a consequence, 2D distributions in the elastic frame can be directly compared with 3D distributions in the Breit frame. Since $\mb \Delta$ is Fourier conjugate to $\mb x$, setting $\Delta^3=0$ amounts to integrating over $x^3$. In other words, the 2D distributions in the elastic frame are just the projections onto the transverse plane of the corresponding 3D distributions in the Breit frame
\begin{equation}
\langle j^z\rangle(b_\perp)=\int\ud x^3\,\langle j^z\rangle(\mb x)\big|_{\mb x=(\mb b_\perp,x^3)}
\end{equation}
with $b_{\perp}=\lvert \mb b_\perp \rvert$ and $j^z=L^z,S^z,J^z_\text{Bel},M^z$, as can be readily checked. \\
Once again, we have
\begin{equation}
\langle J^z\rangle(b_\perp)=\langle L^z\rangle(b_\perp)+\langle S^z\rangle(b_\perp)=\langle J^z_\text{Bel}\rangle(b_\perp)+\langle M^{z}\rangle(b_\perp) \; .\label{kinBel}
\end{equation}
Note also that the dependence on the induced pseudoscalar form factor $G_{P}(t)$ has disappeared because the latter is multiplied by $\Delta^3=0$. \\

Defining the 2D Fourier transform of the form factors as
\begin{equation}
\tilde{F}(b_\perp)=\int\frac{\ud^2 \boldsymbol{\Delta}_\perp}{(2\pi)^2}\,e^{-i\mb\Delta_\perp\cdot\mb b_\perp}\, F(-\mb \Delta^2_\perp) \; , \label{fou2d}
\end{equation}
we can write
\begin{align}
\langle L^{z}\rangle({b}_\perp)&=-\frac{s^z}{2}\,b_{\perp}\,\frac{\ud \tilde{L}(b_{\perp})}{\ud b_{\perp}} \; ,\label{lip}\\
\langle S^{z}\rangle({b}_\perp)&=\frac{s^{z}}{2}\,\tilde{G}_{A}(b_{\perp}) \; ,\\
\langle J^{z}_\text{Bel}\rangle({b}_\perp)&=-\frac{s^z}{2}\,b_{\perp}\,\frac{\ud \tilde{J}(b_{\perp})}{\ud b_{\perp}} \; ,\\
\langle M^{z}\rangle({b}_\perp)&=\frac{s^z}{2}\left[\tilde{G}_{A}(b_{\perp})+\frac{1}{2}\,b_{\perp}\frac{\ud\tilde{G}_A(b_{\perp})}{\ud b_{\perp}}\right] \; .\label{divb}
\end{align}
The impact-parameter density of kinetic total AM $\langle J^{z}\rangle({b}_\perp)=\langle L^{z}\rangle({b}_\perp)+\langle S^{z}\rangle({b}_\perp)$ will differ from the ``naive'' density
\begin{equation}
\langle J^{z}\rangle_{\text{naive}}({b}_\perp)=s^z\tilde{J}(b_{\perp})\label{Lnaive}
\end{equation} 
by a correction term
\begin{equation}
\langle J^{z}\rangle_{\text{corr}}({b}_\perp)=-s^z\left[\tilde{L}(b_{\perp})+\frac{1}{2}\,b_{\perp}\,\frac{\ud \tilde{L}(b_{\perp})}{\ud b_{\perp}}\right] \; .
\label{corrb} \end{equation}
We can also project the 3D monopole and quadrupole contributions to the Belinfante-improved total AM~\eqref{lpol} and \eqref{lquad} onto the transverse plane. This gives
\begin{align}
\langle J^z_\text{Bel}\rangle_{\text{mono}}(b_{\perp})&=\frac{s^{z}}{3}\left[\tilde{J}(b_{\perp})-b_{\perp}\,\frac{\ud\tilde{J}(b_{\perp})}{\ud b_{\perp}}\right] \;,\label{lbur} \\ 
\langle J^z_\text{Bel}\rangle_{\text{quad}}(b_{\perp})&=-\frac{s^{z}}{3}\left[\tilde{J}(b_{\perp})+\frac{1}{2}\,b_{\perp}\,\frac{\ud \tilde{J}(b_{\perp})}{\ud b_{\perp}}\right] \; .\label{lquadrup}
\end{align}
Clearly, the total divergence~\eqref{divb}, the correction~\eqref{corrb} and the quadrupole~\eqref{lquadrup} terms vanish once integrated over $\mb b_\perp$
\begin{equation}
2\pi\int\ud b_\perp\,b_\perp\,\langle M^z\rangle(b_\perp)=2\pi\int\ud b_\perp\,b_\perp\,\langle J^z\rangle_{\text{corr}}(b_{\perp})=2\pi\int\ud b_\perp\,b_\perp\,\langle J^z_\text{Bel}\rangle_{\text{quad}}(b_{\perp})=0,
\end{equation}
as one can see using integration by parts. This explains why the naive $\tilde{J}(b_{\perp})$, the Polyakov-Goeke $\rho^\text{PG}_J(b_\perp)$ and the infinite-momentum frame $\rho^\text{IMF}_J(b_\perp)$ definitions considered by Adhikari and Burkardt~\cite{Adhikari:2016dir} (corresponding in our notations to $\langle J^z\rangle_{\text{naive}}(b_{\perp})$, $\langle J^z_\text{Bel}\rangle_{\text{mono}}(b_{\perp})$ and $\langle J^z_\text{Bel}\rangle(b_{\perp})$, respectively) are different, even though they lead to the same integrated total angular momentum.

\section{Densities in front form}\label{sec4}

As discussed by Burkardt~\cite{Burkardt:2000za}, the density interpretation in the Breit frame is valid only when relativistic effects associated with the motion of the target can be neglected. An elegant way of getting rid of these relativistic corrections is to switch to the front-form dynamics~\cite{Dirac:1949cp,Brodsky:1997de}. In this formalism, the subgroup of Lorentz transformations associated with the transverse plane is Galilean~\cite{Burkardt:2002hr}. As a consequence, there is no need for relativistic corrections as long as we restrict ourselves to the transverse plane.

We introduce light-front coordinates $a^{\mu}=\left[a^+,a^-,\mb{a}_\perp\right]$, with $a^{\pm}=\frac{1}{\sqrt{2}}(a^{0}\pm a^{3})$. Once again we focus on the longitudinal component of angular momentum.
Similarly to the instant-form case, we start with the definition of kinetic OAM distribution in four-dimensional position space
\begin{equation}
\langle L^z\rangle(x)= \varepsilon^{3jk}\,x^j_\perp\int\frac{\ud^{2}\boldsymbol{\Delta}_\perp\,\ud\Delta^+}{(2\pi)^3}\,e^{i\Delta\cdot x}\,\langle T^{+k}\rangle_\text{LF}, \label{l10}
\end{equation}
where
\begin{equation}
\langle T^{\mu\nu}\rangle_\text{LF}\equiv\frac{\langle p',\mb s\lvert T^{\mu\nu}(0)\rvert p,\mb s\rangle}{2\sqrt{p'^+p^+}} . 
\end{equation}
Using the onshell conditions, we can express the light-front energy transfer $\Delta^-$ in terms of the three-momentum transfer $(\Delta^+,\mb\Delta_\perp)$ as
\begin{equation} 
\Delta^-=\frac{\mb{P}_\perp\cdot\mb{\Delta}_\perp-P^-\Delta^+}{P^+},\qquad P^-=\frac{1}{2}\left[\frac{(\mb P_\perp+\frac{\mb\Delta_\perp}{2})^2+M^2}{2(P^++\frac{\Delta^+}{2})}+\frac{(\mb P_\perp-\frac{\mb\Delta_\perp}{2})^2+M^2}{2(P^+-\frac{\Delta^+}{2})}\right].  
\end{equation}
Using integration by parts, and disregarding as usual the surface term, we rewrite Eq.~\eqref{l10} as
\begin{equation} 
\langle L^z\rangle(x)=\varepsilon^{3jk}\int\frac{\ud^{2}\boldsymbol{\Delta}_\perp\,\ud\Delta^+}{(2\pi)^3}\,e^{i\Delta\cdot x}\left[-i\frac{\partial\langle T^{+k}\rangle_\text{LF}}{\partial\Delta^{j}_\perp}+\frac{x^{+}}{2}\left(\frac{p'^j_\perp}{p'^+}+\frac{p^j_\perp}{p^+}\right)\langle T^{+k}\rangle_\text{LF}\right]. \label{LFOAM}
\end{equation}
Densities in the light-front formalism are defined in the Drell-Yan (DY) frame where $\Delta^+= 0$ and $\mb P_\perp=\mb 0_\perp$. This amounts to integrating the four-dimensional distributions over the longitudinal light-front coordinate $x^-$. In such a frame, the dependence on the light-front time $x^+$ in Eq.~\eqref{LFOAM} drops out. The impact-parameter densities of kinetic OAM and spin in the light-front formalism are then given by (see Appendix \ref{a1} for more details)
\begin{align}
\langle L^{z}\rangle(\mb{b}_\perp)&=-i\varepsilon^{3jk}\int\frac{\ud^{2}\boldsymbol{\Delta}_\perp}{(2\pi)^2}\,e^{-i\mb{\Delta}_\perp\cdot\mb{b}_\perp}\left.\frac{\partial\langle T^{+k}\rangle_\text{LF}}{\partial\Delta^{j}_\perp}\right|_\text{DY}\nonumber\\
&=s^z\int\frac{\ud^2\boldsymbol{\Delta}_\perp}{(2\pi)^2}\,e^{-i\mb{\Delta}_\perp\cdot\mb{b}_\perp}\left[L(t)+t\,\frac{\ud L(t)}{\ud t}\right]_{t=-\mb \Delta^2_\perp},\label{L2DLF}\\
\langle S^{z}\rangle(\mb{b}_\perp)&=\frac{1}{2}\,\varepsilon^{3jk}\int\frac{\ud^{2}\boldsymbol{\Delta}_\perp}{(2\pi)^2}\,e^{-i\mb{\Delta}_\perp\cdot\mb{b}_\perp}\left.\langle S^{+jk}\rangle_\text{LF}\right|_\text{DY}\nonumber\\
&=\frac{s^z}{2}\int\frac{\ud^{2}\boldsymbol{\Delta}_\perp}{(2\pi)^2}\,e^{-i\mb{\Delta}_\perp\cdot\mb{b}_\perp}G_{A}(-\mb \Delta^2_\perp) \; .\label{S2DLF}
\end{align}
Similarly, for the impact-parameter densities of Belinfante-improved total angular momentum and total divergence, we find
\begin{align}
\langle J^{z}_\text{Bel}\rangle(\mb{b}_\perp)&=-i\varepsilon^{3jk}\int\frac{\ud^{2}\boldsymbol{\Delta}_\perp}{(2\pi)^2}\,e^{-i\mb{\Delta}_\perp\cdot\mb{b}_\perp}\left.\frac{\partial\langle T^{+k}_\text{Bel}\rangle_\text{LF}}{\partial\Delta^{j}_\perp}\right|_\text{DY}\nonumber\\
&=s^z\int\frac{\ud^2\boldsymbol{\Delta}_\perp}{(2\pi)^2}\,e^{-i\mb{\Delta}_\perp\cdot\mb{b}_\perp}\left[J(t)+t\,\frac{\ud J(t)}{\ud t}\right]_{t=-\mb \Delta^2_\perp}, \label{JBEL2DLF}\\
\langle M^{z}\rangle(\mb{b}_\perp)&=\frac{1}{2}\,\varepsilon^{3jk}\int\frac{\ud^{2}\boldsymbol{\Delta}_\perp}{(2\pi)^2}\,e^{-i\mb{\Delta}_\perp\cdot\mb{b}_\perp}\,\Delta^l_\perp\left.\frac{\partial\langle S^{l+k}\rangle_\text{LF}}{\partial\Delta^j_\perp}\right|_\text{DY}\nonumber\\
&=-\frac{s^z}{2}\int\frac{\ud^{2}\boldsymbol{\Delta}_\perp}{(2\pi)^2}\,e^{-i\mb{\Delta}_\perp\cdot\mb{b}_\perp}\left[t\,\frac{\ud G_{A}(t)}{\ud t}\right]_{t=-\mb \Delta^2_\perp}. \label{M2DLF}
\end{align}
These light-front densities in the Drell-Yan frame coincide with the corresponding instant-form densities in the elastic frame. This should not be too surprising based on the following arguments. Indeed, the Drell-Yan frame is nothing but the elastic frame with $\mb P$ defining the light-front direction. Moreover, instant form and front form coincide in the infinite-momentum frame where $P_z\to\infty$. Since the 2D densities we considered do not depend on $P_z$, they should be the same in both instant form and front form.

\section{Illustration within the scalar diquark model}\label{sec5}

For illustrative purposes, we calculate explicit expressions for the impact-parameter densities in the framework of the scalar-diquark model~\cite{Brodsky:1980zm,Brodsky:1997de,Brodsky:2000xy}. This simple model depicts the nucleon as formed by an active quark and a spectator system described by a scalar diquark.

The quark Light-Front Wave Functions (LFWFs) $\Psi^\Lambda_\lambda(x,\mb k_\perp)$, where $\Lambda=\pm$ and $\lambda=\pm$ denote the helicity of the nucleon and of the quark, respectively, read
\begin{align}
\Psi^+_+(x,\mb k_\perp)&=\Psi^-_-(x,\mb k_\perp)=\left(M+\frac{m}{x}\right)\phi(x,\mb k^2_\perp) ,\\
\Psi^+_-(x,\mb k_\perp)&=-\left[\Psi^-_+(x,\mb k_\perp)\right]^*=-\frac{k^x+ik^y}{x}\,\phi(x,\mb k^2_\perp),
\end{align}
where 
\begin{equation}
\phi(x,\mb k^2_\perp)=-\frac{g\,x\sqrt{1-x}}{\mb k^2_\perp+u(x,m_D^2)}
\end{equation}
and
\begin{equation}
u(x,\mu^2)=x\mu^2+(1-x)m^2-x(1-x)M^2. \label{u}
\end{equation}
Here $g$ is the Yukawa coupling constant, while $m$, $M$ and $m_D$ are the mass of the quark, nucleon and diquark, respectively. 
We define the 2-dimensional Fourier transform of LFWFs from momentum to impact-parameter space as~\cite{Burkardt:2008ua,Kumar:2014coa,Adhikari:2016dir}
\begin{equation}
\Psi^\Lambda_\lambda(x,\mb b_\perp)=\frac{1}{1-x}\int\frac{\ud^{2}\boldsymbol{k}_\perp}{(2\pi)^2}\,e^{i\mb k_\perp\cdot\mb b_\perp/(1-x)}\,\Psi^\Lambda_\lambda(x,\mb k_\perp).
\end{equation}
Writing $\mb b_\perp=b_{\perp}\left(\cos\phi_b,\sin\phi_b\right)$, we obtain
\begin{align}
\Psi^+_+(x,\mb b_\perp)&=\Psi^-_-(x,\mb b_\perp)=-\frac{g\,(xM+m)}{2\pi\,\sqrt{1-x}}\,K_{0}(Z) , \label{lfwf1} \\ 
\Psi^+_-(x,\mb b_\perp)&=\left[\Psi^-_+(x,\mb b_\perp)\right]^*=\frac{ig\,\sqrt{u(x,m_D^2)}\,e^{i\phi_b}}{2\pi\,\sqrt{1-x}}\,K_{1}(Z),\label{lfwf3}
\end{align}
where $K_{n}$ is the $n$-th order modified Bessel function of the second kind and $Z=\sqrt{u(x,m_D^2)}\,b_{\perp}/(1-x)$.

GPDs can be computed using the following LFWF overlap representation in impact-parameter space \cite{Burkardt:2003je}
\begin{align}
\mathcal H(x,b_\perp)&=\frac{1}{2(2\pi)}\left[\lvert\Psi^+_+(x,\mb b_\perp)\rvert^2+\lvert\Psi^+_-(x,\mb b_\perp)\rvert^2\right], \\
-\frac{1}{2M}\left(i\frac{\partial}{\partial b^x}+\frac{\partial}{\partial b^y}\right)\mathcal E(x,b_\perp)&=\frac{1}{2(2\pi)}\left[\Psi^{+*}_+(x,\mb b_\perp)\Psi^-_+(x,\mb b_\perp)+\Psi^{+*}_-(x,\mb b_\perp)\Psi^{-}_{-}(x,\mb b_\perp)\right] , \\
\tilde{\mathcal H}(x,b_\perp)&=\frac{1}{2(2\pi)}\left[\lvert\Psi^+_+(x,\mb b_\perp)\rvert^2-\lvert\Psi^+_-(x,\mb b_\perp)\rvert^2\right] ,
\end{align}
where the Fourier transforms of GPDs are defined as $\mathcal F(x,b_\perp)=\int\frac{\ud^2\boldsymbol{\Delta}_\perp}{(2\pi)^2}\,e^{-i\mb\Delta_\perp\cdot\mb b_\perp}\,F(x,0,-\mb \Delta^2_\perp)$. Using Eqs.~\eqref{lfwf1}-\eqref{lfwf3}, we find
\begin{align}
\mathcal H(x,b_{\perp})&=\frac{g^2}{2(2\pi)^3(1-x)}\left\{(xM+m)^2\left[{K}_{0}(Z)\right]^2+
u(x,m^2_D)\left[{K}_{1}(Z)\right]^2\right\}, \label{Hb}\\
\mathcal E(x,b_{\perp})&=\frac{g^2}{2(2\pi)^3}\,2M(xM+m)\left[{K}_{0}(Z)\right]^2 , \\
\tilde{\mathcal H}(x,b_{\perp})&=\frac{g^2}{2(2\pi)^3(1-x)}\left\{(xM+m)^2\left[{K}_{0}(Z)\right]^2-u(x,m^2_D)\left[{K}_{1}(Z)\right]^2\right\}  . 
\end{align}
Taking the second Mellin moment of these expression, we obtain the EMT form factors in impact-parameter space~\cite{Diehl:2003ny}:
\begin{align}
&\int_{0}^{1}dx\,x\left[\mathcal H(x,b_{\perp})+\mathcal E(x,b_{\perp})\right]=\tilde{A}(b_{\perp})+\tilde{B}(b_{\perp}) , \label{HE}\\
&\int_{0}^{1}dx\, \tilde{\mathcal H}(x,b_{\perp})=\tilde{G}_{A}(b_{\perp})=-\tilde{D}(b_{\perp}) \label{Ga}
\end{align}
which can then be inserted in Eqs~\eqref{lip}-\eqref{lquadrup} to get the various contributions to the density of AM.
\begin{figure}
\centering
\includegraphics[width=0.48\textwidth, keepaspectratio]{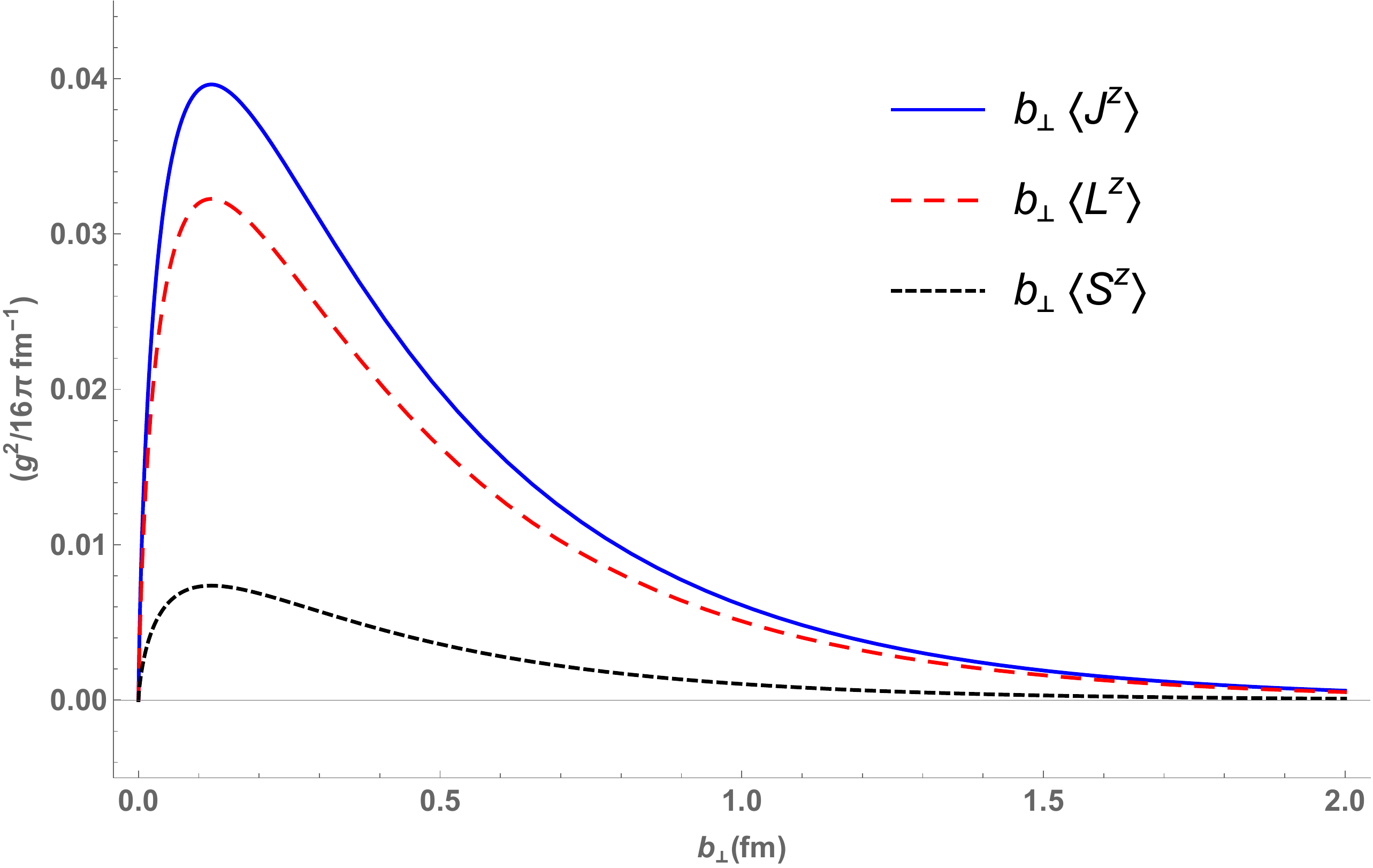}
\hspace{0.3cm}
\includegraphics[width=0.48\textwidth, keepaspectratio]{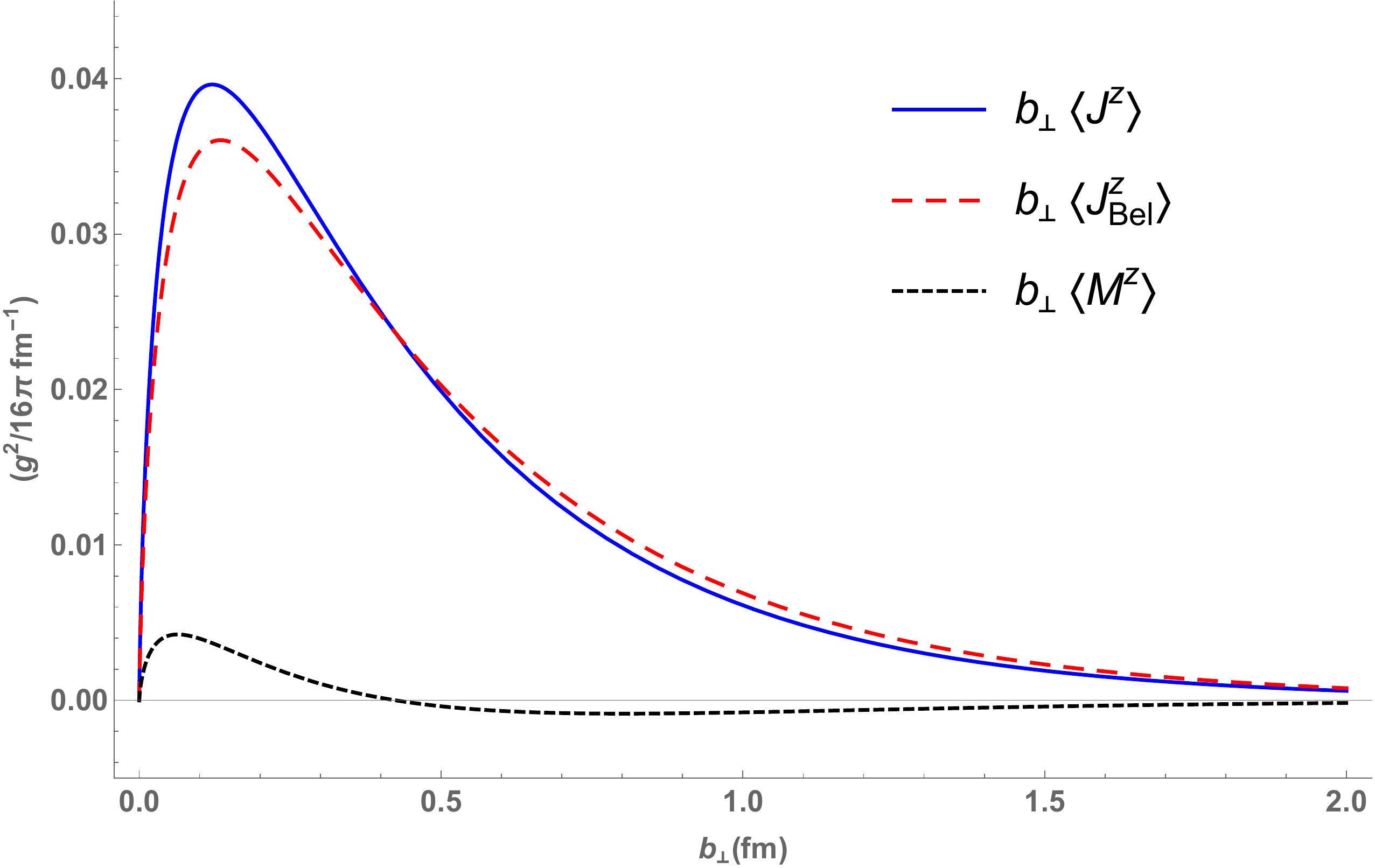} \\
\vspace{0.2cm}
\includegraphics[width=0.48\textwidth, keepaspectratio]{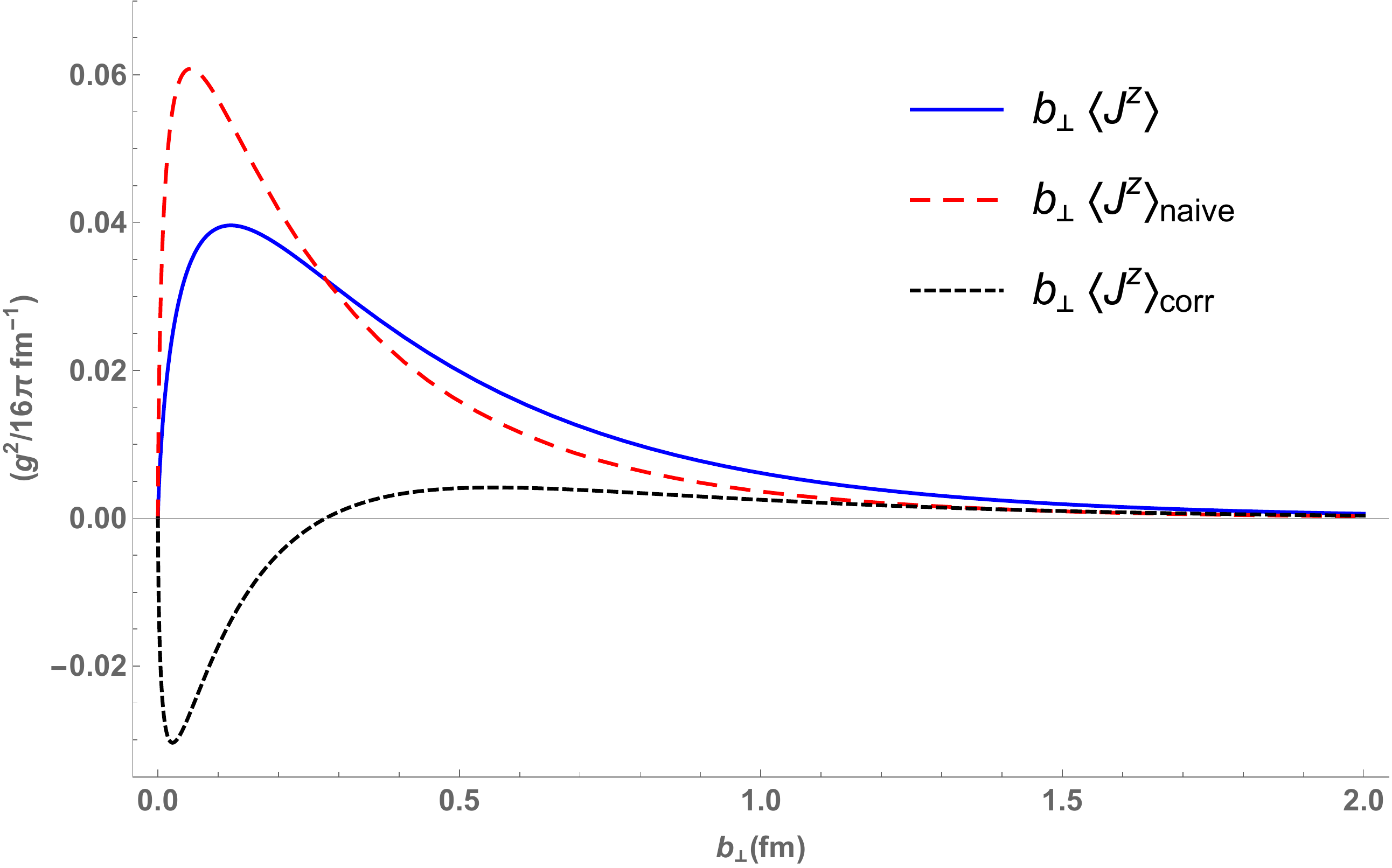}
\hspace{0.3cm}
\includegraphics[width=0.48\textwidth, keepaspectratio]{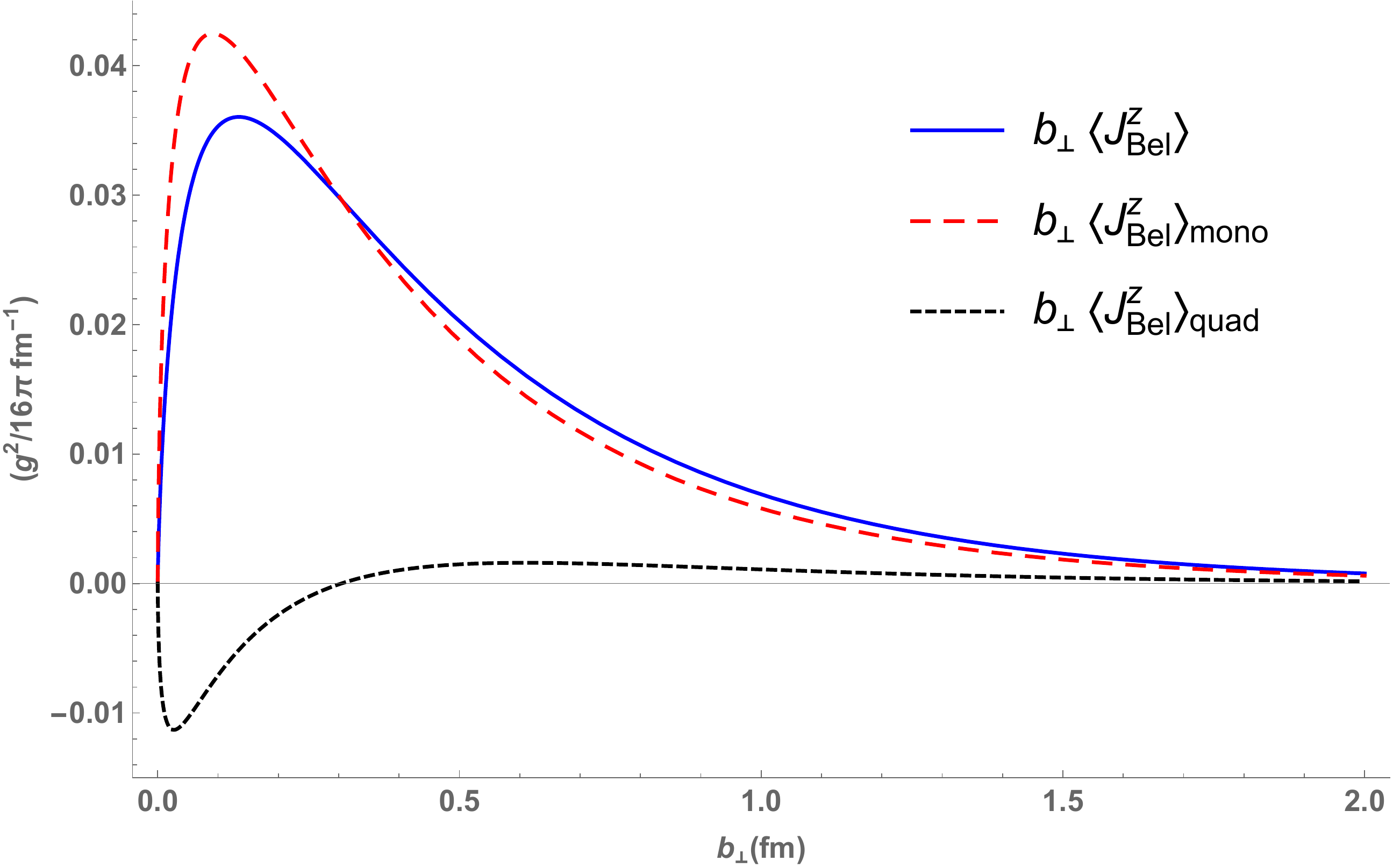}
\caption{\footnotesize{Plots of the density of longitudinal angular momentum in units of $\frac{g^2}{16\pi}$ fm$^{-1}$ as functions of $b_{\perp}=\lvert \mb{b}_\perp\rvert$. All functions are regularized according to Eq.~\eqref{paulivillars}. Upper-left corner: kinetic total AM $\langle J^z\rangle$ (solid line) resulting from the sum of kinetic orbital angular momentum $\langle L^z\rangle$ in Eq.~\eqref{L2DLF} (dashed line) and spin $\langle S^{z}\rangle$ in Eq.~\eqref{S2DLF} (dashed line). Upper-right corner: kinetic total angular momentum $\langle J^z\rangle$ (solid line) expressed as the sum of Belinfante-improved total angular momentum $\langle J^z_\text{Bel}\rangle$ in Eq.~\eqref{JBEL2DLF} (dashed line) and total divergence term $\langle M^{z}\rangle$ in Eq.~\eqref{M2DLF} (dotted line). Lower-left corner: kinetic total angular momentum $\langle J^z\rangle$ (solid line) resulting from the sum of the ``naive'' total angular momentum density $\langle J^z\rangle_{\text{naive}}$ in Eq.~\eqref{Lnaive} (dashed line) and the corresponding correction $\langle J^z\rangle_{\text{corr}}$ in Eq.~\eqref{corrb} (dotted line). Lower-right corner: Belinfante-improved total angular momentum $\left\langle J^z_{\text{Bel}}\right\rangle$ (dashed line) expressed as the sum of the monopole term $\left\langle J^z_{\text{Bel}}\right\rangle_{\text{mono}}$ in Eq.~\eqref{lbur} (dashed line) and the quadrupole contribution $\left\langle J^z_{\text{Bel}}\right\rangle_{\text{quad}}$  in Eq.~\eqref{lquadrup} (dotted line). }
}
\label{plots}
\end{figure}

In Fig.~\ref{plots} we plot the above-mentioned densities as functions of the modulus $b_{\perp}$ of the impact parameter for a longitudinally polarized target $\mb{s}=(0,0,1)$. We choose the same mass parameters as Adhikari and Burkardt~\cite{Adhikari:2016dir}, namely $M=m=m_D=1$ fm$^{-1}$. In order to regulate the ultraviolet divergences $b_{\perp}\rightarrow 0$, we adopt the Pauli-Villars regularization, using the diquark mass $m_D$ as a regulator. More precisely, for each one of the functions $\langle j^z\rangle(b_{\perp};m^2_D)$ considered, we plot
\begin{equation}
b_{\perp}\left[\langle j^z\rangle(b_{\perp};m^2_D)-\langle j^z\rangle(b_{\perp};M^2_D)\right] , \label{paulivillars}
\end{equation}
with $M_D^2=10\,m_D^{2}$. The extra factor of $b_\perp$ comes from the Jacobian of the transformation to polar coordinates.

In the first plot we present the kinetic total AM $\langle J^z\rangle(b_\perp)=\langle L^z\rangle(b_\perp)+\langle S^z\rangle(b_\perp)$ as the sum of kinetic OAM and spin contributions. In the scalar diquark model, both contributions appear to be positive. In the second plot, we compare the kinetic total AM $\langle J^z\rangle(b_\perp)$ with the Belinfante-improved total AM $\langle J^z_\text{Bel}\rangle(b_\perp)$, the difference being attributed to the $\langle M^z\rangle(b_\perp)$ term in Eq.~\eqref{M2DLF}, which originates from  the total-divergence term  in Eq.~\eqref{relation}.
In the third plot, we compare the kinetic total angular momentum $\langle J^z\rangle(b_\perp)$ with the naive density $\tilde{J}(b_{\perp})$. Their difference is given by the correction term $\langle J^z\rangle_{\text{corr}}(b_\perp)$ in Eq.~\eqref{corrb}. In the fourth and last plot, we decomposed the Belinfante-improved total AM  $\langle J^z_\text{Bel}\rangle(b_\perp)=\langle J^z_\text{Bel}\rangle_\text{mono}(b_\perp)+\langle J^z_\text{Bel}\rangle_\text{quad}(b_\perp)$ into its monopole and quadrupole contributions. The monopole contribution is what Adhikari and Burkardt called the Polyakov-Goeke definition~\cite{Adhikari:2016dir}. Once again, although the total divergence term $\langle M^z\rangle(b_\perp)$, the correction term $\langle J^z\rangle_\text{corr}(b_\perp)$ and the quadrupole contribution $\langle J^z_\text{Bel}\rangle_\text{quad}(b_\perp)$ integrate to zero, they need to be taken into account when comparing different definitions for the density of angular momentum.

There is no gauge field in the scalar diquark model we considered. As a consequence, it is expected that the kinetic OAM should coincide with the canonical (or Jaffe-Manohar) OAM~\cite{Leader:2013jra}. The latter can be expressed in terms of the following LFWF overlap representation in impact-parameter space
\begin{equation}
\mathcal{L}^z(b_{\perp})=\frac{1}{2(2\pi)}\int_{0}^{1} \ud x\,(1-x)\,\lvert \Psi^+_-(x,\mb b_\perp)\rvert^2 \; . \label{lfromlfwf}
\end{equation}
Using Eq.~\eqref{lfwf3}, we find
\begin{equation} \label{canexpr}
\mathcal{L}^z(b_{\perp})=\frac{g^2}{2(2\pi)^3}\int_{0}^{1} \ud x\,u(x,m^2_D)\left[{K}_{1}(Z)\right]^2 \; .
\end{equation} 
Note that the canonical OAM can alternatively be defined in terms of generalized transverse momentum parton distributions~\cite{Lorce:2011kd,Lorce:2011ni}, leading to the same expression as in Eq.~\eqref{lfromlfwf} in the scalar diquark model~\cite{Kanazawa:2014nha}.
This  has to be compared with the expression for the kinetic OAM $\langle L^z\rangle(b_\perp)$ that we obtain from Eq.~\eqref{lip}, using~\eqref{Hb}-\eqref{Ga}:
\begin{equation}\label{kinexpr}
\begin{aligned}
\langle L^z\rangle(b_\perp)&=\frac{g^2}{2(2\pi)^3}\,\frac{1}{2}\int_0^1\ud x\,\frac{1}{1-x}\,\Big\{\left[(1-x)(x^2M^2-m^2)+(1+x)\,u(x,m_D^2)\right]Z K_0(Z)K_1(Z)\\
&\qquad\qquad\qquad\qquad\qquad\qquad +(1+x)\,u(x,m_D^2)\left[{K}_{1}(Z)\right]^2\Big\}.
\end{aligned}
\end{equation}
Using integration by parts, one can show that $\langle L^z\rangle(b_\perp)=\mathcal L^z(b_\perp)$ for $b_\perp>0$, see Appendix \ref{a2}. To the best of our knowledge, this is the first time that the equality between kinetic and canonical OAM is checked explicitly at the density level. We also understand the failure to observe the equality in Ref.~\cite{Adhikari:2016dir} as coming from the fact that the authors incorrectly defined the density of kinetic OAM as
\begin{equation}
L^z_\text{IMF}(b_\perp)\equiv\langle J^z_\text{Bel}\rangle(b_\perp)-\left\langle S^{z}\right\rangle(b_\perp)
\end{equation} 
which misses the total divergence term $\langle M^z\rangle(b_\perp)$ as one can see from Eq.~\eqref{kinBel}.

\section{Conclusions}\label{sec6}

In this work, we addressed the question of the definition of angular momentum at the density level.  One often makes use of the freedom offered by superpotentials to deal with a symmetric energy-momentum tensor, as motivated by General Relativity. In the context of Particle Physics, however, spin densities play a fundamental role and make the energy-momentum tensor asymmetric. In particular, we showed that for a spin-$1/2$ target the form factor accounting for the antisymmetric part of the energy-momentum tensor coincides (up to a sign) with the axial-vector form factor. This provides an interesting new way of calculating the latter on the lattice. While superpotential terms do not play any role at the level of integrated quantities, it is of crucial importance to keep track of them at the density level.

We revisited Polyakov's work on the three-dimensional distribution of angular momentum in the Breit frame. Working with an asymmetric energy-momentum tensor allowed us to derive directly the correct density of orbital angular momentum. Densities in the Breit frame can be extended to the more general class of elastic frame, provided one projects onto a two-dimensional plane. Thanks to this generalization, we were able to establish a simple connection between instant-form densities defined in the Breit frame and light-front densities defined in the Drell-Yan frame for the longitudinal components of angular momentum. 

We used the scalar diquark model to illustrate our results. We showed explicitly that when all the terms integrating to zero are included in the expressions, no discrepancies are found between the different definitions of angular momentum. In particular, we checked for the first time explicitly that the canonical and kinetic angular momentum do coincide at the density level, as expected in a system without gauge bosons.

\section*{Acknowledgments}

This work was partially supported by the European Research Council (ERC) under the European Union's Horizon 2020 research and innovation programme (grant agreement No. 647981, 3DSPIN).

\appendix

\section{Dirac bilinears}\label{a1}

In this Appendix, we collect the Dirac bilinears involved in the calculation of the various matrix elements of \cref{sec3} and \cref{sec4}.\\

\noindent In the Breit frame, we used
\begin{align}
\overline u\left(\tfrac{\bm\Delta}{2},\bm s\right)\gamma_5 u\left(-\tfrac{\bm\Delta}{2},\bm s\right)&=-(\bm \Delta\cdot\bm s),\\
\overline u\left(\tfrac{\bm\Delta}{2},\bm s\right)\gamma^k\gamma_5 u\left(-\tfrac{\bm\Delta}{2},\bm s\right)&=2P^0s^k-\frac{\Delta^k(\bm\Delta\cdot\bm s)}{2(P^0+M)},\\
\overline u\left(\tfrac{\bm\Delta}{2},\bm s\right)i\sigma^{k\lambda}\Delta_\lambda u\left(-\tfrac{\bm\Delta}{2},\bm s\right)&=-2M\,i\epsilon^{klm}\Delta^ls^m.
\end{align}
In the elastic frame, we used
\begin{align}
\overline u\left(P_z,\tfrac{\bm\Delta_\perp}{2},\bm s\right)\gamma^3\gamma_5 u(P_z,-\tfrac{\bm\Delta_\perp}{2},\bm s)&=2P^0\,s^z,\\
\overline u\left(P_z,\tfrac{\bm\Delta_\perp}{2},\bm s\right)i\sigma^{k\lambda}\Delta_\lambda u(P_z,-\tfrac{\bm\Delta_\perp}{2},\bm s)&=-2M\,i\epsilon^{kl3}\Delta^ls^z,\qquad k=1,2.
\end{align}
In the Drell-Yan frame with light-front spinors, we used
\begin{align}
\overline u_\text{LF}\left(P^+,\tfrac{\bm\Delta_\perp}{2},\bm s\right)\gamma^+\gamma_5 u_\text{LF}(P^+,-\tfrac{\bm\Delta_\perp}{2},\bm s)&=2P^+s^z,\\
\overline u_\text{LF}\left(P^+,\tfrac{\bm\Delta_\perp}{2},\bm s\right)i\sigma^{k\lambda}\Delta_\lambda u_\text{LF}\left(P^+,-\tfrac{\bm\Delta_\perp}{2},\bm s\right)&=-2M\,i\epsilon^{kl3}\Delta_{\perp}^ls^z,\qquad k=1,2.
\end{align}

\section{Kinetic and canonical orbital angular momentum}\label{a2}

Proving the equality between Eqs.~\eqref{canexpr} and \eqref{kinexpr} amounts to establishing the following identity
\begin{equation}\label{cankinidentity}
\int_0^1\ud x\,\frac{1}{1-x}\left[(1-x)(x^2M^2-m^2)+(1+x)\,u\right]Z K_0(Z)K_1(Z)=\int\ud x\,\frac{1-3x}{1-x}\,u\left[K_1(Z)\right]^2 \; ,
\end{equation}
Using
\begin{equation}
\frac{1}{Z}\,\frac{\partial Z}{\partial x}=\frac{1}{2u}\,\frac{\partial u}{\partial x}+\frac{1}{1-x},\qquad x\,\frac{\partial u}{\partial x}=u+x^2M^2-m^2,
\end{equation}
we find that
\begin{equation}
\frac{1}{1-x}\left[(1-x)(x^2M^2-m^2)+(1+x)\,u\right]Z=2ux\,\frac{\partial Z}{\partial x}.
\end{equation}
Noting now that
\begin{equation}
\frac{\ud(Z^2\left[K_1(Z)\right]^2)}{\ud Z}=-2Z^2K_0(Z)K_1(Z),
\end{equation}
we can rewrite the LHS of Eq.~\eqref{cankinidentity} as
\begin{equation}
\int_0^1\ud x\,\frac{1}{1-x}\left[(1-x)(x^2M^2-m^2)+(1+x)\,u\right]Z K_0(Z)K_1(Z)=-\int_0^1\frac{xu}{Z^2}\,\frac{\partial(Z^2\left[K_1(Z)\right]^2)}{\partial x}.
\end{equation}
Integrating by parts, the boundary term vanishes identically for $b_\perp>0$ and we obtain the RHS of Eq.~\eqref{cankinidentity}.

\bibliographystyle{myrevtex}
\bibliography{EMT_biblio}

\end{document}